\documentclass[prb,twocolumn,aps,superscriptaddress,showpacs]{revtex4}
\usepackage{amsmath}
\usepackage{graphicx}
\usepackage{latexsym}
\usepackage{amsfonts}
\usepackage{amssymb}

\begin{document}

\def\tg{\mbox{\textsl{g}}}
\def\bS{\mbox{\boldmath $\Sigma$}}
\def\bDelta{\mbox{\boldmath $\Delta$}}
\def\bcalE{\mbox{\boldmath ${\cal E}$}}
\def\bcalG{\mbox{\boldmath ${\cal G}$}}
\def\bG{\mbox{\boldmath $\Gamma$}}
\def\bSS{\mbox{\boldmath $S$}}
\def\bK{\mbox{\boldmath $K$}}
\def\bT{\mbox{\boldmath $T$}}
\def\bU{\mbox{\boldmath $U$}}
\def\bV{\mbox{\boldmath $V$}}
\def\bH{\mbox{\boldmath $H$}}
\def\bQ{\mbox{\boldmath $Q$}}
\def\bme{\mbox{\boldmath $e$}}
\def\bg{\mbox{\boldmath $g$}}
\def\bq{\mbox{\boldmath $q$}}
\def\bp{\mbox{\boldmath $p$}}
\def\bh{\mbox{\boldmath $h$}}
\def\gC{\mbox{\boldmath $C$}}
\def\gZ{\mbox{\boldmath $Z$}}
\def\gR{\mbox{\boldmath $R$}}
\def\gN{\mbox{\boldmath $N$}}
\def\gM{\mbox{\boldmath $M$}}
\def\gO{\mbox{\boldmath $O$}}
\def\ua{\uparrow}
\def\da{\downarrow}
\def\a{\alpha}
\def\b{\beta}
\def\g{\gamma}
\def\G{\Gamma}
\def\d{\delta}
\def\D{\Delta}
\def\e{\epsilon}
\def\ve{\varepsilon}
\def\z{\zeta}
\def\h{\eta}
\def\th{\theta}
\def\k{\kappa}
\def\l{\lambda}
\def\L{\Lambda}
\def\m{\mu}
\def\n{\nu}
\def\x{\xi}
\def\X{\Xi}
\def\p{\pi}
\def\P{\Pi}
\def\r{\rho}
\def\s{\sigma}
\def\S{\Sigma}
\def\t{\tau}
\def\f{\phi}
\def\vf{\varphi}
\def\F{\Phi}
\def\c{\chi}
\def\w{\omega}
\def\W{\Omega}
\def\Q{\Psi}
\def\q{\psi}
\def\de{\partial}
\def\inf{\infty}
\def\ra{\to}
\def\bra{\langle}
\def\ket{\rangle}

\title{Time-dependent quantum transport: A practical scheme using density 
functional theory}

\author{S. Kurth}
\affiliation{Institut f\"ur Theoretische Physik, Freie Universit\"at Berlin, 
Arnimallee 14, D-14195 Berlin, Germany}

\author {G. Stefanucci}
\affiliation{Solid State Theory, Institute of Physics, Lund University, 
S\"olvegatan 14 A, S-22362 Lund, Sweden}

\author{C.-O. Almbladh}
\affiliation{Solid State Theory, Institute of Physics, Lund University, 
S\"olvegatan 14 A, S-22362 Lund, Sweden}

\author{A. Rubio}
\affiliation{Departamento de F\'{i}sica de Materiales, Facultad de Ciencias 
Qu\'{i}micas, UPV/EHU, Unidad de Materiales Centro Mixto CSIC-UPV/EHU 
and Donostia International  Physics Center (DIPC), San Sebasti\'{a}n, Spain}

\author{E.K.U. Gross}
\affiliation{Institut f\"ur Theoretische Physik, Freie Universit\"at Berlin, 
Arnimallee 14, D-14195 Berlin, Germany}

\date{\today}

\begin{abstract}

We present a computationally tractable scheme of time-dependent 
transport phenomena within open-boundary 
time-dependent density-functional-theory. Within this approach all the 
response properties of a system are determined from 
the time-propagation of the set of ``occupied'' Kohn-Sham orbitals
under the influence of the external bias. This central idea is combined
with an open-boundary description of the  geometry of the system that is
divided into three regions: left/right leads and 
the device region (``real simulation region''). 
We  have derived a general scheme to extract the set of initial states in
the device region that will
be propagated in time with proper transparent boundary-condition at the 
device/lead interface. This is possible due to a new modified Crank-Nicholson 
algorithm that allows an efficient time-propagation of open quantum systems. 
We illustrate the method in one-dimensional model systems as a first step 
towards a full first-principles implementation. In particular we show how
a stationary current develops in the system independent of the 
transient-current history upon application of the bias. The present work 
is ideally suited to study ac transport and photon-induced charge-injection. 
Although the implementation has been done assuming clamped ions, we discuss 
how it can be extended to include dissipation due to electron-phonon coupling 
through the combined simulation of the electron-ion dynamics as well as 
electron-electron correlations.

\end{abstract}


\maketitle

\section{Introduction}

During the last decades, the size of electronic circuits has continuously 
been reduced. Today, systems like quantum wires and quantum dots are routinely 
produced on the nanometer scale. Recently, the seemingly ultimate limit of 
minituarization has been achieved by several experimental groups who were 
able to place single molecules between two macroscopic electrodes.
\cite{ReedZhouMullerBurginTour:97,
SansDevoretDaiThessSmalleyGeerligsDekker:97}
To describe transport properties on such a small scale, a quantum theory 
of transport is required.\cite{jortner:97,book}

A cornerstone of such a theory is the Landauer-B\"uttiker formalism 
\cite{Landauer:57,Buettiker:86} which provides a method to compute the 
steady-state current of non-interacting electrons for meso- or nanoscopic 
systems connecting two (or more) macroscopic electrodes.

Alternatively, the technique of nonequilibrium Green functions (NEG)
\cite{KadanoffBaym:62,Keldysh:65} 
has been used to tackle quantum transport. Studies using the NEG 
approach typically use tight-binding-like model Hamiltonians to describe 
the combined system electrodes plus ``device''. 
A well known scheme is the one introduced by Caroli {\em et al.}
\cite{caroli1,caroli2} In the remote past the left and 
right electrodes are disconnected and in equilibrium at two different 
chemical potentials; the conducting part of the Hamiltonian is switched 
on adiabatically and eventually a steady-state develops. 
Within this contacting approach 
also time-dependent transport phenomena have been investigated.
\cite{JauhoWingreenMeir:94}
Caroli {\em et al.} discussed non-interacting systems only. Their 
approach has later been extended to account for short-range 
electron-electron interaction and for interaction with vibrations in 
the device region.\cite{MeirWingreen:92} An excellent overview 
of the field can be 
found in the book by Haug and Jauho \cite{HaugJauho:98} and in 
Ref.~\onlinecite{book}. Despite its appeal, 
the above scheme has limitations since the 
time-dependent perturbation is the tunneling Hamiltonian, connecting 
the electrodes to the device, rather than the external electric 
field. 

Cini proposed another scheme \cite{cini} also based on NEG. Here, 
the system electrodes plus ``device'' is 
connected and in equilibrium in the remote past.
The time-dependent perturbation is the external scalar potential. 
It has been shown \cite{saprb04} that for non-interacting systems the 
contacting approach and the Cini approach   
yield the same current in the long-time limit and that in the dc case the 
steady state current does not depend on the history of the applied potential.
Moreover, the Cini scheme is well suited for a density functional 
extension since the electrons are driven out of equilibrium by a local 
potential rather than by a non-local one (see below).

With recent experimental progress to place single molecules as devices between 
macroscopic electrodes there also has been considerable activity to describe 
transport through these systems on an {\em ab initio} level. 
Most approaches are based on a self-consistency procedure first 
proposed by Lang.\cite{lang} In this steady-state approach based on density 
functional theory (DFT), exchange and correlation is approximated by the 
static Kohn-Sham (KS) potential and the charge density is obtained 
self-consistently in the presence of the steady current.
However, the original justification 
involved subtle points such as different 
Fermi levels deep inside the left and right electrodes and the 
implicit reference of non-local perturbations such as tunneling 
Hamiltonians within a DFT framework. (For a 
detailed discussion we refer to Ref. \onlinecite{saprb04}.) 
The steady-state DFT approach has been further developed 
\cite{DerosaSeminario:01,BrandbygeMozosOrdejonTaylorStokbro:02,XueDattaRatner:02,Calzolari:04} 
and the results have been most useful for understanding the 
qualitative behavior of measured current-voltage characteristics. Quantitatively, 
however, the theoretical I-V curves often differ from the experimental 
ones by several orders of magnitude.\cite{DiVentraPantelidesLang:00} 
Several explanations are possible for such a mismatch: 
models are not sufficiently refined, 
parasitic effects in measurements have been underestimated, 
the characteristics of the molecule-contact interface 
are not well understood and difficult to address given their atomistic complexity.
Adding to the theoretical reason for this discrepancy is the fact that the 
transmission functions computed from static DFT have resonances at the 
non-interacting Kohn-Sham excitation energies which in general do not 
coincide with the true excitation energies. Furthermore, different 
exchange-correlation functionals lead to DFT-currents that vary by 
more than an order of magnitude.\cite{Krstic:03}

Excitation energies of interacting systems are accessible via time-dependent 
(TD) DFT.\cite{RungeGross:84,PetersilkaGossmannGross:96} In 
this theory, the time-dependent density of an interacting system moving in 
an external, time-dependent local potential can be calculated via a fictitious 
system of non-interacting electrons moving in a local, effective 
time-dependent potential. Therefore this theory is in principle well suited 
for the treatment of nonequilibrium transport problems.\cite{StefanucciAlmbladh:04} 
A basic issue is that most exchange-correlation functionals
have been derived under equilibrium conditions and their application to 
non-equilibrium problems should be analyzed in more detail. However, this is beyond
the scope of the present work.

Before a TDDFT calculation of transport can be tackled, a number of 
technical problems have to be addressed. In particular, one needs a practical 
scheme for the propagation of the time-dependent Schr\"odinger equation for 
an infinitely large system. Of course, since one can in practice only deal 
with finite systems this can only be achieved by applying the correct 
boundary conditions. The problem of so-called ``transparent boundary 
conditions'' for the time-dependent Schr\"odinger equation has been 
attacked by many authors. For a recent overview, the reader is referred to 
Ref.~\onlinecite{Moyer:04}. 

In this paper we present a propagation scheme which is particularly designed 
to be used for the calculation of time-dependent transport problems. In 
Section \ref{secII}, we combine the Cini scheme with TDDFT and we 
develop a general formalism based on the 
propagation of Kohn-Sham orbitals in open systems. 
In Section \ref{secIII} we will address the question of how 
to obtain the correct initial states for the propagation.
An algorithm for the time-evolution of open systems 
is proposed in Section \ref{secIV}. It is based on a 
modified version of the Crank-Nicholson algorithm.  
Section~\ref{details} describes some details of our numerical 
implementation and Section~\ref{examples} gives results 
for several one-dimensional model systems. 
We draw our conclusions in Section \ref{conc}.

\section{General Formulation}
\label{secII}

\begin{figure}[t]
\includegraphics[scale=0.9]{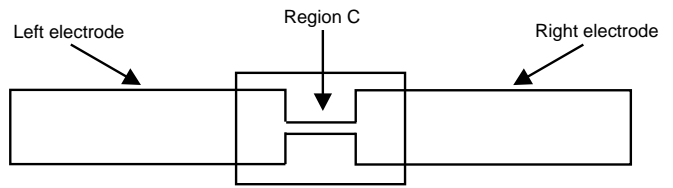}
\caption{\label{system} Schematic diagram of the system described in the 
main text for the calculation of charge transport through the central
constriction.}
\label{curr1}
\end{figure}

We consider an electrode-junction-electrode system which is initially 
in equilibrium ($t<0$). The system is contacted and no current flows through 
the junction, the charge density of the electrodes being perfectly 
balanced, see Fig. \ref{system}. 
Therefore, the system initially is in its ground state which, due to the 
Hohenberg-Kohn theorem,\cite{HohenbergKohn:64} is a functional of the 
density. This density can then be computed in the usual way by 
$n({\bf r},0)=\sum_{occ} |\q_{s}({\bf r},0)|^{2}$ where the sum is over 
the occupied Kohn-Sham orbitals $\q_{s}({\bf r},0)$, i.e., the eigenfunctions 
of the Kohn-Sham Hamiltonian $\bH(0)$ with eigenergy below the Fermi energy. 
Here and in the following we use boldface notation to denote operators
in one-electron Hilbert space.

To observe a current we drive the system out of equilibrium by 
exposing the electrons to an external electric potential (bias). 
Without loss of generality we will assume that this  
external bias vanishes for times $t\leq0$. According to the Runge-Gross 
theorem \cite{RungeGross:84}, the time-dependent density can be 
calculated by evolving the KS orbitals according to  
the KS equation of TDDFT $i\dot{\q}_{s}(t)=\bH(t)\q_{s}(t)$ where $\bH(t)$ is 
the time-dependent KS Hamiltonian. Thus, 
$n({\bf r},t)=\sum_{occ}|\q_{s}({\bf r},t)|^{2}$ and the 
continuity equation is  
$\frac{\partial}{\partial t}n({\bf r},t)=
-\nabla\cdot{\bf j}_{\rm KS}({\bf r},t)$, 
where ${\bf j}_{\rm KS}({\bf r},t)=-\sum_{occ}{\rm Im}[
\q_{s}^{*}({\bf r},t)\nabla\q_{s}({\bf r},t)]$ is the KS current 
density. Due to the Runge-Gross theorem and the continuity equation one can 
deduce that the longitudinal part of the KS current density equals the 
longitudinal part of the true current density. This need not be true 
for the transverse part of the current density. However, the transverse 
part of the current density does not contribute to the total current which can 
then be calculated by a surface integral
\begin{equation}
I_{S}(t)=e\sum_{s=1}^{N}\int_{S}{\rm d}\s\; \hat{{\bf n}}\cdot{\rm Im}[
\q_{s}^{*}({\bf r},t)\nabla\q_{s}({\bf r},t)],
\label{curror}
\end{equation}
where $\hat{{\bf n}}$ is the unit vector perpendicular to the surface 
element ${\rm d}\s$ and the surface $S$ is perpendicular to the longitudinal 
geometry of our system. In order to propagate the KS 
orbitals we need to solve the Schr\"odinger equation for a macroscopic 
and non-periodic system. This goal is hopeless unless we know 
the dynamics of the remote parts of the system. 
We restrict ourselves to metallic electrodes. Then, the external 
potential and the disturbance introduced by the device region are 
screened deep inside the electrodes.
As the system size increases, the remote parts are less disturbed by the 
junction and the density inside the electrodes approaches  
the equilibrium bulk-density. Thus, the macroscopic size of the 
electrodes leads to an enormous simplification since the initial-state  
self-consistency is not disturbed far away from the constriction. 

It is convenient to partition the system into three main regions: 
a central region $C$ consisting of the junction and a few atomic layers of the left 
and right electrodes and two regions $L$, $R$ which describe the left 
and right bulk electrodes. Only the central device region $C$ will be 
treated explicitly. Our scheme accounts for the full dynamical 
screening in the central region. It can be further refined by taking 
into account screening effects also deeper in the electrodes at the 
level of linear response, with a limited increase in numerical 
efforts. (These effects might be of importance in the initial transient 
phase where long-range plasma oscillations in the electrodes may 
occur.) 
According to the above partitioning, the original 
KS Hamiltonian can be written as a $3\times 3$ block matrix, and 
the Schr\"odinger equation reads 
\begin{equation}
i\frac{\partial}{\partial t}\left[\begin{array}{c}
\q_{L} \\ \q_{C} \\ \q_{R}
\end{array}
\right]=
\left[
\begin{array}{ccc}
\bH_{LL} & \bH_{LC} & 0 \\
\bH_{CL} & \bH_{CC} & \bH_{CR} \\
0 & \bH_{RC} & \bH_{RR}
\end{array}
\right]
\left[\begin{array}{c}
\q_{L} \\ \q_{C} \\ \q_{R}
\end{array}
\right],
\label{tdse}
\end{equation}
where $\q_{\a}$ is the projected wave-function onto the region $\a=L,R,C$. 
We can solve the diffential equation for $\q_{L}$ and $\q_{R}$ by 
introducing the retarded Green function $\bg_{\a}$ for electrode $\a$, which 
satisfies $\left\{
i\frac{{\rm d}}{{\rm d} t}-\bH_{\a\a}(t)
\right\}\bg_{\a}(t,t')=\d(t-t')$, $\a=L,R$,
with boundary conditions $\bg_{\a}(t^{+},t)=-i$ and $\bg_{\a}(t,t^{+})=0$. 
Then, we have for $\a=L,R$
\begin{equation}
\q_{\a}(t)=i\bg_{\a}(t,0)\q_{\a}(0)+\int_{0}^{t}{\rm d}t'\bg_{\a}(t,t')
\bH_{\a C}\q_{C}(t').
\label{leadeq}
\end{equation}
Using Eq. (\ref{leadeq}), the equation for $\q_{C}$ can be written as 
\begin{eqnarray}
i\frac{\partial}{\partial t}\q_{C}(t)&=&
\bH_{CC}(t)\q_{C}(t)+\int_{0}^{t}{\rm d}t'\bS(t,t')\q_{C}(t')
\nonumber \\ 
&& +i\sum_{\a=L,R}\bH_{C\a}
\bg_{\a}(t,0)\q_{\a}(0),
\label{evolc}
\end{eqnarray}
where $\bS=\sum_{\a=L,R}\bH_{C\a}\bg_{\a}\bH_{\a C}$ is the 
self-energy which accounts for the hopping in and out of region $C$.
Thus, for any given KS orbital we can evolve its projection 
onto the central region by solving Eq.~(\ref{evolc}) 
in region $C$. Eq.~(\ref{evolc}) has also been derived elsewhere 
(for static Hamiltonians).\cite{HellumsFrensley:94} 
To summarize, all the
complexity of the infinite electrode-junction-electrode system shown in
Fig.~\ref{curr1} has been reduced to the solution of an 
open quantum-mechanical system (the central region $C$) with proper 
time-dependent boundary conditions. 

\section{Computation of extended eigenstates}
\label{secIII}

Eq.~(\ref{evolc}) of the previous Section is the central equation of our 
approach to time-dependent transport. It is a reformulation of 
the original time-dependent Schr\"odinger equation (\ref{tdse}) of the full 
system in terms of an equation for the central (device) region only. The coupling 
to the leads is taken into account by the lead Green functions $\bg_{\a}, 
\a = L,R$. Eq.~(\ref{evolc}) has the structure of a time-dependent 
Schr\"odinger equation with two extra terms: the term involving the 
self energy $\bS$ which we will also denote as the memory integral 
since it involves the wavefunction in the central region at previous 
times during the propagation.~\cite{memory} 
The second term describes the injection of particles induced by a 
non-vanishing projection of the initial wave-function onto the leads.

Eq.~(\ref{evolc}) is first order in time, therefore we need to specify an 
initial state which is to be propagated. We want to study the time 
evolution of systems perturbed out of their equilibrium ground state. 
Of course, the ground state of our noninteracting system is the Slater 
determinant of the occupied eigenstates of the full, extended Hamiltonian 
in equilibrium, $\bH(t<0) = \bH^s$. 
The practical question then arises how one can obtain these eigenstates 
without having to deal explicitly with the extended Hamiltonian. Note 
that, unlike in a bulk solid, the translational symmetry is broken in 
the present device situation.

In the present Section we propose a solution of this problem which is based 
on the partitioning approach used in many steady-state transport calculations 
(see, {\em e.g.}, Ref. \onlinecite{Datta:95}). The retarded Green function of 
the static Hamiltonian in the energy domain is determined by
\begin{equation}
\left[ (E + i \eta) \mathbf{1} - \bH^s \right] \bcalG(E) = \mathbf{1} \; .
\label{g-energy}
\end{equation} 
The Green function $\bcalG(E)$ of the full system can be written 
in the same block structure as the Hamiltonian
\begin{equation}
\bcalG(E) = \left[
\begin{array}{ccc}
\bcalG_{LL}(E) & \bcalG_{LC}(E) & \bcalG_{LR}(E) \\
\bcalG_{CL}(E) & \bcalG_{CC}(E) & \bcalG_{CR}(E) \\
\bcalG_{RL}(E) & \bcalG_{RC}(E) & \bcalG_{RR}(E) 
\end{array}
\right].
\end{equation}
Eq.~(\ref{g-energy}) can be solved for the block of the Green function 
referring only to the central region 
\begin{eqnarray}
\lefteqn{ \bcalG_{CC}(E) = } \nonumber \\
&& \!\!\!\!\! \frac{1}{ (E+ i \eta) \mathbf{1}_C - \bH_{CC}^s 
- \sum_{\a=L,R} \bH_{C\a}^s \bg_{\a}(E) \bH_{\a C}^s }
\label{gcc}
\end{eqnarray}
with the retarded Green function of lead $\a$  
\begin{equation}
\bg_{\a}(E) = \frac{1}{(E+i \eta) \mathbf{1}_{\a} - \bH_{\a\a}^s}
\label{glead}
\end{equation}
and the unit matrix $\mathbf{1}_{\a}$ in region $\a$. This Green function 
enters as a central ingredient into the Fisher-Lee relation 
\cite{FisherLee:81} for the calculation of the transmission function. 
Through the coupling to the leads it provides for level broadening of the 
isolated central part, but it also contains information on the eigenstates of 
the extended system.

In order to illustrate the central idea of our method  
to extract the extended eigenstates from the 
Green function we consider $\bH^s$ to be the discretized form of a 
continuous Hamiltonian $\hat{H}^s({\bf r})$. 
The continous Green function and the discretized one for uniform lattice 
spacing $\Delta x$ are connected by 
\begin{equation}
{\cal G}({\bf r}_i,{\bf r}_j,E) = \bcalG(E)_{ij}/(\Delta x)^N
\label{g-contdis}
\end{equation}
where $N=1,2,3$ is the number of spatial dimensions of the problem. 
We choose the convention that a single-particle orbital 
$\psi_{Ej}$ of the Hamiltonian $\hat{H}^s$ is uniquely specified by 
its eigenenergy $E$ and a label $j$ for the $d_E$ degenerate orbitals 
of this energy. 
Using the Lehmann representation and assuming that  $\hat{H}^s$ is 
invariant under time-reversal, the imaginary part of ${\cal G}$ is
\begin{equation}
-\frac{1}{\pi} {\rm Im} \,\left[ {\cal G}({\bf r},{\bf r}',E) \right] =  
\sum_{E'}\d(E-E')
\sum_{j=1}^{d_{E'}} \psi_{E'j}({\bf r}) 
\psi_{E'j}^*({\bf r}') \; .
\label{img-cont}
\end{equation}
Multiplying Eq.~(\ref{img-cont}) by $\psi_{El}({\bf r}')$, 
integrating over ${\bf r}'$, using the orthogonality of the single particle
states and dividing by the density of states 
$D(E)=\sum_{E'}\d(E-E')d_{E'}$ we obtain
\begin{equation}
-\frac{1}{\pi D(E)} \int \! \! {\rm d}^N r \; {\rm Im} \, 
\left[{\cal G}({\bf r},{\bf r}',E) \right] \psi_{El}({\bf r}') =\g(E) \psi_{El}({\bf r})
\label{img-eigen}
\end{equation}
with
\begin{equation}
\g(E)=\frac{1}{D(E)}\sum_{E'}\d(E-E').
\end{equation}
Eq.~(\ref{img-eigen}) has the structure of an eigenvalue equation where  
the energy eigenstate $\psi_{El}$ is also an eigenstate of the 
integral operator $-{\rm Im}\,[{\cal G}({\bf r},{\bf r}',E)]/(\pi D(E))$ with eigenvalue 
$\g(E)$. For a given energy $E$, this integral operator has $d_E$ different 
degenerate eigenstates. 

We note that Eq.~(\ref{img-cont}) is valid for {\em all} 
points in space, in particular also for both ${\bf r}$ and ${\bf r}'$ 
representing points in the central region. In this case we know the 
Green function $\bcalG_{CC}$ through Eqs.~(\ref{gcc}) and (\ref{g-contdis}). 
Below we show that the 
eigenfunctions of ${\rm Im}[\bcalG_{CC}]$ 
can be expressed as linear combination of the $\psi_{El}$. 
Let us consider the matrix formed by the elements
\begin{equation}
b_{ml} = \int_C \! \! {\rm d}^N r \; \psi_{Em}^{*}({\bf r}) \psi_{El}({\bf r}) \; .
\end{equation}
where the integration is over the central region only. This matrix 
is Hermitian and can be diagonalized, {\em i.e.}, 
\begin{equation}
\sum_{l=1}^{d_E} b_{ml} a_{jl} = \lambda_j a_{jm}
\end{equation}
with $\lambda_j$ real. Next we compute the matrix elements of the Green 
function with respect to the functions 
\begin{equation}
a_{Ej}({\bf r}) = \sum_{l=1}^{d_E} a_{jl}(E) \psi_{El}({\bf r}) \; .
\end{equation}
After a straightforward manipulation one finds
\begin{eqnarray}
-\frac{1}{\pi} \int_C \! \! {\rm d}^N r \; \int_C \! \! {\rm d}^N r' \; 
a_{El}({\bf r}) {\rm Im} \, [{\cal G}({\bf r},{\bf r}',E)] a_{Ej}({\bf r}') \nonumber \\
= \delta_{jl} \lambda_j^2 \sum_{E'} \delta(E-E') 
\end{eqnarray}
which shows explicitly that the functions $a_{Ej}({\bf r})$
diagonalize ${\rm Im} \,[\bcalG_{CC}]$ 
in the central region and that the eigenvalues are positive. Since 
any linear combination of degenerate eigenstates is again an eigenstate, 
diagonalizing ${\rm Im}\,[\bcalG_{CC}(E)]$ gives us one set of linearly independent, 
degenerate eigenstates of energy $E$. 

In our practical implementation described in more detail in Section 
\ref{details}, we diagonalize
\begin{equation}
-\frac{1}{\pi D_{C}(E)} {\rm Im}\,[\bcalG_{CC}(E)]
\label{img-pract}
\end{equation}
where
\begin{equation}
D_C(E) = -\frac{1}{\pi} {\rm Tr} \, \left\{{\rm Im}\,[\bcalG_{CC}(E)]\right\}
\end{equation}
is the total density of states in the central region. If we use $N_g$ grid 
points to describe the central region, the diagonalization in principle 
gives $N_g$ eigenvectors but only a few have the physical meaning of 
extended eigenstates at this energy. It is, however, very easy to identify 
the physical states by looking at the eigenvalues: only few eigenvalues 
(for the simple examples we studied either one or two)
are nonvanishing  and they always add up to unity. 
The corresponding states are the physical 
ones. All the other eigenvalues are zero (or numerically close to zero) and 
the corresponding states have no physical meaning. 

The procedure described above gives the correct extended eigenstates only up 
to a normalization factor. When diagonalizing Eq. (\ref{img-pract}) 
with typical library routines one obtains eigenvectors which are normalized 
to the central region. Physically this 
might be incorrect. Therefore, the normalization has to be 
fixed separately. In the example of Section \ref{details} we fixed the norm 
by matching the wavefunction for the central region to the known form 
(and normalization) of the 
wavefunction in the macroscopic leads. 

It should be emphasized that the procedure described here for the extraction 
of eigenstates of the extended system from $\bcalG_{CC}(E)$ only works in practice 
if $E$ is in the continuous part of the spectrum due to the sharp peak 
of the delta function in the discrete part of the spectrum. 

Eigenstates in the discrete part of the spectrum can be found by 
considering the original Schr\"odinger equation for the full system:
\begin{equation}
\bH^s \psi = E \psi.
\end{equation}
Using again the block structure of the Hamiltonian this can be transformed 
into an effective Schr\"odinger equation for an {\em energy-dependent} Hamiltonian 
for the central region only:
\begin{equation}
\left( \bH^s_{CC} - \sum_{\alpha = L,R} \bH_{C\a} \bg_{\alpha}(E) 
\bH_{\alpha C} \right) \psi_C = E \psi_C.
\end{equation}
This equation has solutions only for certain values of $E$ which are 
the discrete eigenenergies of the full Hamiltonian $\hat{H}^{s}$.
Therefore, one can find these states 
by iteration. We have succesfully tested this idea for systems where 
the analytic solutions are known. However, since the main focus of the 
present work is transport where the continuum states are the essential 
physical ingredient, we will not deal with the states in the discrete 
spectrum for the remainder of this paper. Those states might play a role
in the description of charge-accumulation in molecular transport,
as, {\em e.g.}, in Coulomb blockade phenomena.

\section{Algorithm for time evolution}
\label{secIV}

In order to calculate the longitudinal current in an 
electrode-junction-electrode system we need to propagate the 
Kohn-Sham orbitals. The main difficulty stems from the macroscopic 
size of the electrodes whose remote parts,ultimately, are taken 
infinitely far away from the central, explicitly treated, scattering 
region $C$.

The problem can be solved by imposing transparent boundary 
conditions\cite{Moyer:04} at the electrode-junction interfaces. 
Efficient algorithms have been  
proposed for wave-packets initially {\em localized} in the scattering 
region and for Hamiltonians constant in time. In this Section 
we propose an algorithm well suited for  
delocalized initial states, as well as for localized ones, evolving 
with a time-dependent Hamiltonian. 

Let $\bH(t)$ be the time-dependent KS Hamiltonian. 
We partition $\bH(t)$ as  
in Section \ref{secII}. The explicitly treated region $C$ includes  
the first few atomic layers of the left and right electrodes. The boundaries of 
this region are chosen in such a way that the density outside $C$ 
is accurately described by an equilibrium bulk density. 
It is convenient to write $\bH_{\a\a}(t)$, with $\a=L,R$, as the sum of a 
term $\bH^{s}_{\a\a}$ which is constant in time and another term $\bU_{\a}(t)$ 
which is explicitly time-dependent, $\bH_{\a\a}(t)=\bH^{s}_{\a\a}+\bU_{\a}(t)$. 
In configuration space $\bU_{\a}(t)$ 
is diagonal at any time $t$ since the KS potential is local in space.
Furthermore, the diagonal elements $U_{\a}({\bf r},t)$ are spatially 
constant for metallic electrodes. Thus, 
$\bU_{\a}(t)=U_{\a}(t){\bf 1}_{\a}$ and $U_{L}(t)-U_{R}(t)$ is the 
total potential drop across the central region. 
We write $\bH(t)=\tilde{\bH}(t)+\bU(t)$ with 
\begin{equation}
\tilde{\bH}(t)=\left[
\begin{array}{ccc}
\bH^{s}_{LL} & \bH_{LC} & 0 \\
\bH_{CL} & \bH_{CC}(t) & \bH_{CR} \\
0 & \bH_{RC} & \bH^{s}_{RR}
\end{array}
\right],
\end{equation}
and
\begin{equation}
\bU(t)=\left[
\begin{array}{ccc}
U_{L}(t){\bf 1}_{L} & 0 & 0 \\
0 & 0 & 0 \\
0 & 0 & U_{R}(t){\bf 1}_{R}
\end{array}
\right].
\end{equation}
In this way, the only term  
in $\tilde{\bH}(t)$ that depends on $t$ is $\bH_{CC}(t)$. 
For any given initial state $\q(0)=\q^{(0)}$ we calculate 
$\q(t_{m}=m\D t)=\q^{(m)}$ by using a generalized form of 
the Cayley method 
\begin{widetext}
\begin{equation}
(1+i\d 
\tilde{\bH}^{(m)})\frac{1+i\frac{\d}{2}\bU^{(m)}}{1-i\frac{\d}{2}\bU^{(m)}}\q^{(m+1)}=
(1-i\d 
\tilde{\bH}^{(m)})\frac{1-i\frac{\d}{2}\bU^{(m)}}{1+i\frac{\d}{2}\bU^{(m)}}\q^{(m)},
\label{prop}
\end{equation}
\end{widetext}
with 
$\tilde{\bH}^{(m)}=\frac{1}{2}[\tilde{\bH}(t_{m+1})+\tilde{\bH}(t_{m})]$, 
$\bU^{(m)}=\frac{1}{2}[\bU(t_{m+1})+\bU(t_{m})]$ and $\d=\D t/2$. 
It should be noted that our propagator is 
norm conserving (unitary) and 
accurate to second-order in $\d$, as is the Cayley propagator.~\cite{castro} 
Denoting by $\q_{\a}$ the projected wave function 
onto the region $\a=R,L,C$, we find from Eq.~(\ref{prop}) 
\begin{equation}
\q_{C}^{(m+1)}=
\frac{1-i\d \bH_{\rm eff}^{(m)}}{1+i\d \bH_{\rm eff}^{(m)}}
\q_{C}^{(m)}
+S^{(m)}-M^{(m)}.
\end{equation}
Here, $\bH_{\rm eff}^{(m)}$ is the effective Hamiltonian of the central region:
$\bH_{\rm eff}^{(m)}=\bH_{CC}^{(m)}-
i\d\bH_{CL}(1+i\d\bH^{s}_{LL})^{-1}\bH_{LC}
-i\d\bH_{CR}(1+i\d\bH^{s}_{RR})^{-1}\bH_{RC}$ with
$\bH_{CC}^{(m)}=\frac{1}{2}[\bH_{CC}(t_{m+1})+\bH_{CC}(t_{m})]$. The source term 
$S^{(m)}$ describes the injection of density into the region $C$, 
while the memory term $M^{(m)}$ is responsible for the hopping in and 
out of the region $C$. In terms of the propagator for the uncontacted 
and undisturbed $\a$ electrode
\begin{equation}
\bg_{\a}=\frac{1-i\d\bH^{s}_{\a\a}}{1+i\d\bH^{s}_{\a\a}},
\label{discgreen}
\end{equation}
the source term can be written as 
\begin{equation}
S^{(m)}=-\frac{2i\d}{1+i\d \bH_{\rm eff}^{(m)}} 
\sum_{\a=L,R}\frac{\L_{\a}^{(m,0)}}{u_{\a}^{(m)}}\bH_{C\a}
\frac{[\bg_{\a}]^{m}}{1+i\d\bH^{s}_{\a\a}}\q_{\a}^{(0)},
\end{equation}
with
\begin{equation}
u_{\a}^{(m)}=\frac{1-i\frac{\d}{2}U_{\a}^{(m)}}{1+i\frac{\d}{2}U_{\a}^{(m)}}
\quad {\rm and}\quad
\L_{\a}^{(m,k)}=\prod_{j=k}^{m}[u_{\a}^{(j)}]^{2}.
\end{equation}
For a wave packet initially localized in $C$ the projection onto the left 
and right electrode $\q_{\a}^{(0)}$ vanishes and $S^{(m)}=0$ for any 
$m$, as it should be. The memory term is more 
complicated and reads
\begin{widetext}
\begin{equation}
M^{(m)}=-\frac{\d^{2}}{1+i\d \bH_{\rm eff}^{(m)}}\sum_{\a=L,R}\sum_{k=0}^{m-1}
\frac{\L_{\a}^{(m,k)}}{u_{\a}^{(m)}u_{\a}^{(k)}}
[\bQ_{\a}^{(m-k)}+\bQ_{\a}^{(m-k-1)}]
(\q_{C}^{(k+1)}+\q_{C}^{(k)})
\end{equation}
\end{widetext}
where
\begin{equation}
\bQ_{\a}^{(m)}=\bH_{C\a}\frac{[\bg_{\a}]^{m}}{1+i\d\bH^{s}_{\a\a}}\bH_{\a C}.
\label{Qdef}
\end{equation}
The quantities $\bQ_{\a}^{(m)}$ depend on the geometry of the 
system and are independent of the initial state $\q^{(0)}$. 

\begin{figure}[b]
\includegraphics[scale=0.55]{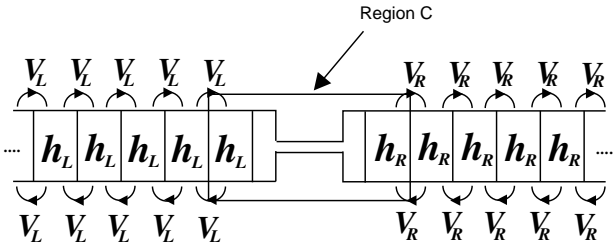}
\caption{\label{semi} Schematic sketch of an electrode-junction-electrode 
system with semiperiodic electrodes.}
\end{figure}

Below we propose a recursive scheme to calculate the $\bQ_{\a}^{(m)}$'s for 
those system geometries having semiperiodic electrodes along the 
longitudinal direction, see Fig. \ref{semi}. In this case $\bH^{s}_{\a\a}$ 
has a tridiagonal block form
\begin{equation}
\bH^{s}_{\a\a}=\left[
\begin{array}{cccc}
    \bh_{\a} & \bV_{\a} & 0 & \ldots \\
    \bV_{\a} & \bh_{\a} & \bV_{\a} & \ldots \\ 
    0 & \bV_{\a} & \bh_{\a} & \ldots \\ 
    \ldots & \ldots & \ldots & \ldots
\end{array}    
\right],
\end{equation}
where $\bh_{\a}$ describes a convenient cell and $\bV_{\a}$ is the 
hopping Hamiltonian between two nearest neighbor cells. 
Without loss of generality we assume that both $\bh_{\a}$ and $\bV_{\a}$ 
are square matrices of dimension $N_{\a}\times N_{\a}$. 
Taking into account that the central region 
contains the first few cells of the left and right electrodes, the 
matrix $\bQ_{\a}^{(m)}$ has the following structure
\begin{equation}
\bQ_{L}^{(m)}=\left[
\begin{array}{ccc}
\bq_{L}^{(m)} & 0 & 0 \\
0 & 0  & 0 \\
0 & 0 & 0
\end{array}    
\right],\quad
\bQ_{R}^{(m)}=\left[
\begin{array}{ccc}
0 & 0 & 0 \\
0 & 0  & 0 \\
0 & 0 & \bq_{R}^{(m)}
\end{array}    
\right].
\end{equation}
The $\bq_{\a}^{(m)}$'s are square matrices of dimension $N_{\a}\times 
N_{\a}$ and are given by
\begin{equation}
\bq_{\a}^{(m)}=\bV_{\a}\left[
\frac{[\bg_{\a}]^{m}}{1+i\d\bH_{\a\a}}
\right]_{1,1}\bV_{\a},
\end{equation}
where the subscript $(1,1)$ denotes the first diagonal block of the 
matrix in the square brackets. We introduce the generating matrix function
\begin{equation}
\bq_{\a}(x,y)\equiv\bV_{\a}\left[
\frac{1}{x+iy\d\bH_{\a\a}}
\right]_{1,1}\bV_{\a},
\label{q1def}
\end{equation}
which can also be expressed in terms of continued matrix fractions (see Appendix 
\ref{app})
\begin{equation}
\bq_{\a}(x,y)=\bV_{\a}\frac{1}{x+iy\d\bh_{\a}+y^{2}\d^{2}\bq_{\a}(x,y)}\bV_{\a}.
\label{qalpha}
\end{equation}
The $\bq_{\a}^{(m)}$'s can be obtained from
\begin{equation}
\bq_{\a}^{(m)}=\frac{1}{m!}\left.\left[-\frac{\de}{\de x}+\frac{\de }{\de y}\right]^{m}
\bq_{\a}(x,y)\right|_{x=y=1}.
\label{gen}
\end{equation}

From Eqs.~(\ref{gen}) and (\ref{qalpha}) one can build up a recursive 
scheme. Let us define 
$\bp_{\a}^{-1}(x,y)=x+iy\d\bh_{\a}+y^{2}\d^{2}\bq_{\a}(x,y)$ and 
$\bp_{\a}^{(m)}=\frac{1}{m!}[-\frac{\de}{\de x}+\frac{\de}{\de y}]
\bp_{\a}(x,y)|_{x=y=1}$. Then, by definition,
\begin{equation}
\bq_{\a}^{(m)}=\bV_{\a}\bp_{\a}^{(m)}\bV_{\a}.
\end{equation}
Using the identity $\frac{1}{m!}[-\frac{\de}{\de x}+\frac{\de}{\de y}]^{m}
\bp_{\a}(x,y)\bp_{\a}^{-1}(x,y)=0$, one finds
\begin{eqnarray}
(1+i\d\bh_{\a})\bp_{\a}^{(m)}=(1-i\d\bh_{\a})\bp_{\a}^{(m-1)}
\quad\quad\quad\quad\quad
\nonumber \\ -
\d^{2}\sum_{k=0}^{m}(\bq_{\a}^{(k)}+2\bq_{\a}^{(k-1)}+\bq_{\a}^{(k-2)})
\bp_{\a}^{(m-k)}
\label{recur}
\end{eqnarray}
with $\bp_{\a}^{(m)}=\bq_{\a}^{(m)}=0$ for $m<0$. Once 
$\bq_{\a}^{(0)}$ has been obtained by solving Eq.~(\ref{qalpha}) 
with $x=y=1$, we can calculate $\bp_{\a}^{(0)}=
[1+i\d\bh_{\a}+\d^{2}\bq_{\a}^{(0)}]^{-1}$. Afterwards, we can use 
Eq.~(\ref{recur}) with $\bq_{\a}^{(1)}=\bV_{\a}\bp_{\a}^{(1)}\bV_{\a}$ 
to calculate $\bp_{\a}^{(1)}$ and hence $\bq_{\a}^{(1)}$ and so on and 
so forth. 

This concludes the description of our algorithm for the propagation of the 
time-dependent Schr\"odinger equation for extended systems. It is worth  
mentioning an additional complication here which arises for the propagation 
of a time-dependent Kohn-Sham equation. This complication stems from the fact 
that in order to compute $\q_{C}^{(m+1)}$ at time step $m+1$ one needs to 
know the time-dependent KS potential at the same time step which, via the 
Hartree and exchange-correlation potentials, depends on the yet unknown 
orbitals $\q_{C}^{(m+1)}$. Of course, the solution is to use a 
predictor-corrector approach: in the first step one approximates 
$\bH_{CC}^{(m)}$ by $\bH_{CC}(t_m)$, computes new orbitals 
$\tilde{\q}_{C}^{(m+1)}$ and from those obtains an improved approximation 
for $\bH_{CC}^{(m)}$. 

\section{Implementation Details}
\label{details}

All the methodological discussion above
 is general and can be applied to general device 
configurations as long as they can be mapped into a longitudinal-like geometry
as described in Fig.~\ref{semi}.  In order to demonstrate the feasibility 
of the scheme described in the previous 
Sections we have implemented it for one-dimensional model systems. The 
extension to real molecular-device configurations is presently under 
development.~\cite{wirtz} 
We have used a simple three-point discretization for 
the second derivative 
\begin{equation}
\frac{{\rm  d}^2}{{\rm d} x^2} \psi(x) \vert_{x=x_i} \approx 
\frac{1}{(\Delta x)^2} \left( \psi(x_{i+1}) - 2 \psi(x_i) + 
\psi(x_{i-1}) \right)
\end{equation}
with equidistant grid points $x_i$, $i=1,\dots,N_g$ and spacing $\Delta x$. 
Within this approximation matrices of the form $\bH_{C\a} \gM \bH_{\a C}$ 
which are $N_g \times N_g$ matrices and appear, {\em e.g.}, in Eq.~(\ref{gcc}) 
or (\ref{Qdef}), have only one nonvanishing matrix element. For $\a=L$ this 
is the $(1,1)$ element, for $\a=R$ it is the $(N_g,N_g)$ element. 

In order to proceed we have to specify the nature of the leads and therefore
the lead Green function. Here we choose the simplest case of 
semi-infinite, uniform leads at constant potential $U_{\a 0}$. In this case, 
the Green function (\ref{glead}) in the energy domain can be given in 
closed form:
\begin{eqnarray}
\lefteqn{
[\bg_{\a}(E)]_{kl} = - \frac{i \Delta x}{\sqrt{2 \tilde{E}_{\a}}} 
\exp\left\{i \sqrt{2 \tilde{E}_{\a}} | x_k - x_l|\right\}} \nonumber \\
&+& \frac{i \Delta x}{\sqrt{2 \tilde{E}_{\a}}} 
\exp\left\{i \sqrt{2 \tilde{E}_{\a}}( | x_k - x_{\a0}| + |x_l - 
x_{\a0}|)\right\}
\end{eqnarray}
with $\tilde{E}_{\a} = E-U_{\a 0}$. The abscissa $x_{\a 0}$ 
is the position of the 
interface between the lead and the device region and $x_k = x_{\a 0} 
\pm k \Delta x$, where the plus sign applies for $\a=R$ and the minus sign 
for $\a=L$.

\begin{figure}[t]
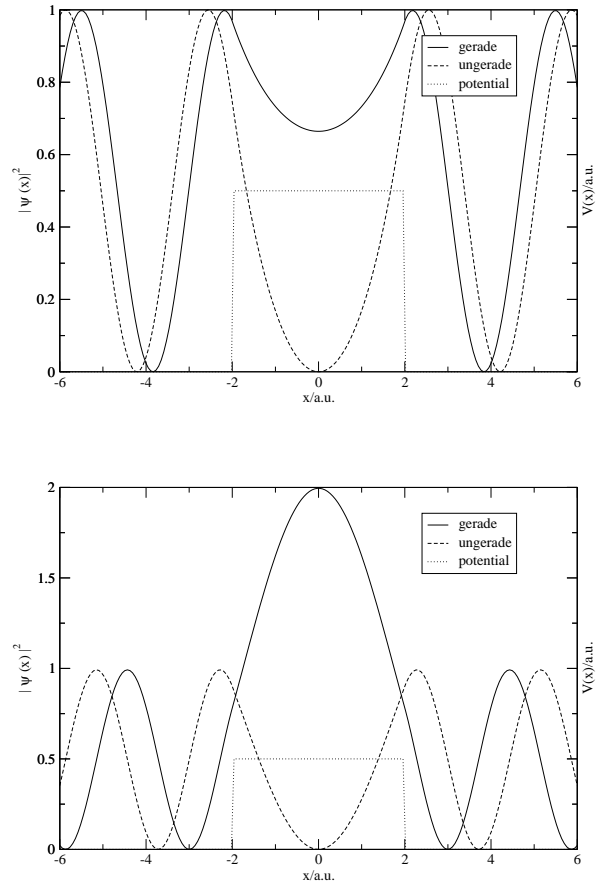

\includegraphics[width=0.9\columnwidth]{fig3up.eps}\\[1.0cm]
\includegraphics[width=0.9\columnwidth]{fig3low.eps}\\[0.5cm]
\caption{\label{ext-states} Continuum states of square potential barrier 
at different energies with leads at zero potential. Upper panel: 
eigenstates for $\ve=0.45$ a.u., just below the barrier height of $0.5$ a.u.. 
Lower panel: eigenstates at $\ve=0.6$ a.u..}
\end{figure}

The results of the procedure for calculating extended eigenstates as described 
in Section \ref{secIII} is illustrated in Fig.~\ref{ext-states} for a 
square potential barrier with zero potential in both leads. 
In the upper panel we 
have the square modulus of eigenstates at an energy below the barrier height 
while in the lower panel eigenstates with energy higher than the barrier 
are shown. The states result from diagonalization of Eq.~(\ref{img-pract}). 
In order to obtain the normalization constant we compute the logarithmic 
derivative at the boundary of the central region numerically and match it 
to the analytic form in the lead to obtain the phase shift $\delta_{\a}$:
\begin{equation}
\frac{1}{2} \frac{{\rm d}^2}{{\rm d} x^2} \ln( |\psi(x)|^2 ) 
\bigg\vert_{x=x_{\a 0}} = q \cot(\delta_{\a})
\end{equation}
where $q= \sqrt{2 \tilde{E}_{\a}}$. Knowing the phase shift we can rescale the 
wavefunction such that it matches with the analytic form 
$\sin(q (x-x_{\a 0}) + \delta_{\a})$ at the interface. Of course, this form 
of the extended states only applies for $\tilde{E}_{\a} >0$ but as long as 
$E$ is in the continuous part of the spectrum, it is correct at least for 
one of the leads. This is sufficient to determine the normalization. The 
states obtained numerically with this procedure coincide with the known 
analytical results. 

We then implemented the propagation scheme presented in 
the previous Section. Within  our three-point approximation, 
$\bh_{\a}$, $\bV_{\a}$ and $\bq_{\a}$ are 
$1\times 1$ matrices. The equation for $q_{\a}^{(0)}$ [see 
Eqs.~(\ref{qalpha}) and (\ref{gen})] becomes a simple quadratic equation 
which can be solved explicitly
\begin{equation}
q_{\a}^{(0)} = \frac{- (1 + i \delta h_{\a}) 
+ \sqrt{(1 + i \delta h_{\a}) ^2 + 4 (\delta V_{\a})^2}}
{2 \delta^2}  \, .
\label{q0}
\end{equation}
Although the quadratic equation has two solutions, the above choice for 
$q_{\a}^{(0)}$ is dictated by the fact that the Taylor expansions for small 
$\delta$ of Eqs.~(\ref{q0}) and (\ref{qalpha}) have to coincide. 
Using this result we then 
solved the iterative scheme to obtain the $q_{\a}^{(m)}$ for $m \geq 1$. 

\begin{figure}[t]
\includegraphics[width=0.9\columnwidth]{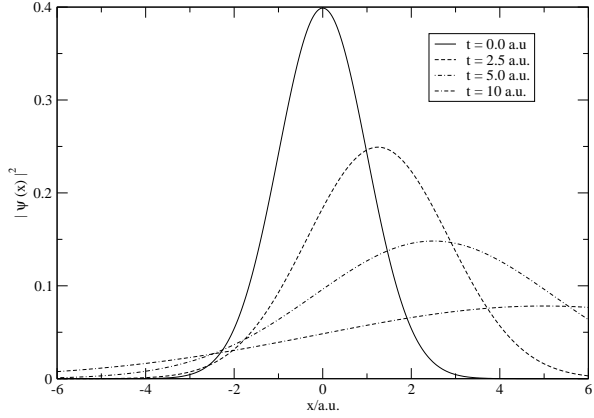}
\caption{\label{wavepacket} Time evolution of a Gaussian wavepacket with 
initial width 1.0 a.u. and initial momentum 0.5 a.u. for various 
propagation times. The transparent boundary conditions allow the wavepacket 
to pass the propagation region without spurious reflections at the boundaries.}
\end{figure}

As a first check on the propagation method we propagated a Gaussian 
wavepacket which, at initial time $t=0$, is completely localized in the 
central device region. (The source terms $S^{(m)}$ then vanish identically). 
As can be seen in Fig.~\ref{wavepacket}, the wavepacket correctly propagates 
through the boundaries without any spurious reflections.

For the propagation of the extended initial states (the eigenstates of the 
unperturbed system) we also need to implement the source terms $S^{(m)}$. In 
the following we assume that the left and right leads are at the same 
potential initially so that the analytic form of the initial states 
is in both leads given by $\sin(q(x-x_{\a 0}) + \delta_{\a}) = 
\cos(\delta_{\a}) \sin(q(x-x_{\a 0})) + \sin(\delta_{\a}) 
\cos(q(x-x_{\a 0}))$. The propagation of the term proportional to 
$\sin(q(x-x_{\a 0}))$ is trivial since this is an eigenstate of the lead 
Hamiltonian with energy $\varepsilon_q=q^2/2$. 
Therefore, if - in discretized form - $\psi_{R,1}^{(0)} = 
(\sin(q \Delta x), \sin(2 q \Delta x), \ldots )^T$ we obtain:
\begin{equation}
\bH_{CR}\frac{[\bg_R]^m}{1 + i \delta \bH_{RR}^s} \psi_{R,1}^{(0)} = 
V_{R} \frac{(1 - i \delta \varepsilon_q)^m}
{(1 + i \delta \varepsilon_q)^{m+1}}\, \bme_R
\end{equation}
and similarly for the left lead. Here, 
$\bH_{RR}^s$ is the static part of the right-lead Hamiltonian, 
$\bg_R$ the corresponding Green function according to definition 
(\ref{discgreen}) and $\bme_R=(0,\ldots,0,1)^{T}$ is a 
unit vector.

\begin{figure}[t]
\includegraphics[width=0.9\columnwidth]{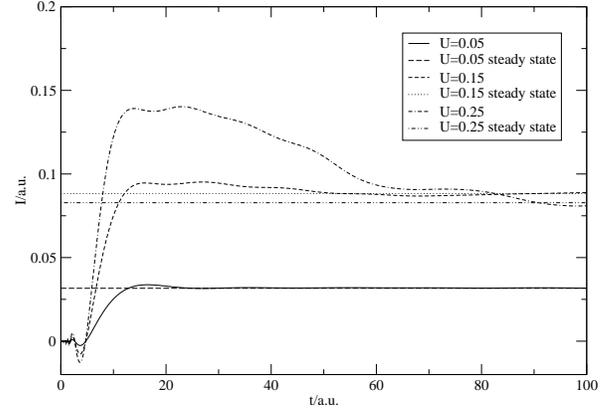}
\caption{\label{examp1} Time evolution of the current for a system where  
initially the potential is zero in the leads and the propagation 
region. At $t=0$, a constant bias with opposite sign in the left and right 
leads is switched on, $U=U_L=-U_R$ (values in atomic units). 
The propagation region extends from 
$x=-6$ to $x=+6$ a.u.. The Fermi energy of the initial state is $\ve_F=0.3$ 
a.u.. The current in the center of the propagation region is shown.}
\end{figure}

The propagation of the part proportional to $\cos(q(x-x_{\a 0}))$ is 
more complicated since this is not an eigenstate of the lead Hamiltonian. 
We define the function $R(x,y)$ from
\begin{equation}
R(x,y) \bme_R = \bH_{CR} \frac{1}{x + i y \delta \bH_{RR}^s} \psi_{R,2}^{(0)} 
\; ,
\end{equation}
where $\psi_{R,2}^{(0)} = (\cos(q \Delta x), \cos(2 q \Delta x), \ldots )^T$.  
Introducing the tridiagonal matrix 
\begin{equation}
\gO = \left( 
\begin{array}{ccccc}
0 & 1 & 0 & 0 & \cdots \\
1 & 0 & 1 & 0 & \cdots \\
0 & 1 & 0 & 1 & \cdots \\
\vdots & \vdots & \vdots & \vdots & \ddots 
\end{array} \right)
\end{equation}
and using 
\begin{equation}
V_R \gO \psi_{R,2}^{(0)} = (\ve_q - h_R) \psi_{R,2}^{(0)} - V_R 
(1,0,0,\ldots)^T 
\end{equation}
one arrives at
\begin{equation}
R(x,y) = \frac{1}{x + i y \delta \ve_q} \left[ V_R \cos(q \Delta x) + 
i y \delta  q_R(x,y) \right],
\end{equation}
where $q_R(x,y)$ is given by Eq.~(\ref{q1def}). Defining 
\begin{equation}
R^{(m)} = \frac{1}{m!} \left(- \frac{\partial}{\partial x} + 
\frac{\partial}{\partial y} \right)^m R(x,y) \big\vert_{x=y=1} 
\end{equation}
one finds
\begin{eqnarray}
\lefteqn{
R^{(m)} = \frac{(1- i \delta \ve_q)^m}{(1 + i \delta \ve_q)^{m+1}} V_R 
\cos(q \Delta x)} \nonumber \\
&& + i \delta  \sum_{k=0}^{m} \frac{(1- i \delta \ve_q)^{m-k}}
{(1+ i \delta \ve_q)^{m+1-k}} \left(q_R^{(k)} + q_R^{(k-1)} \right)
\end{eqnarray}
and finally
\begin{equation}
\bH_{CR}\frac{[\bg_R]^m}{1 + i \delta \bH_{RR}^s} \psi_{R,2}^{(0)} = 
R^{(m)} \bme_R.
\end{equation}
One may proceeds along the same lines for extracting the left 
component.

To test our implementation we have propagated eigenstates of the extended 
system. As expected, these states just pick up an exponential phase factor 
$\exp(-i E t)$ during the propagation.

\section{Examples}
\label{examples}

We are now in a position to apply our algorithm to the calculation of 
time-dependent currents in one-dimensional model systems. The systems 
are initially in thermodynamic equilibrium. At time $t=0$, 
a bias is switched on in the electrodes. 

As a first example we considered a system where the electrostatic potential 
vanishes identically both in the left and right leads as well as in the 
central region which is explicitly propagated. Initially, all single particle 
levels are occupied up to the Fermi energy $\ve_F$. At $t=0$ a 
constant bias is switched on in the leads and the time-evolution of the 
system is calculated. We chose the bias in the right lead as the negative 
of the bias in the left lead, $U_R = - U_L$. The current is 
calculated from Eq. (\ref{curror}):
\begin{eqnarray}
\lefteqn{
I(x,t) = 2 \int_{-k_F}^{k_F} \frac{{\rm d} k}{2 \pi} \;\;
{\rm Im}\, \left( \psi_k^*(x,t) \frac{{\rm d}}{{\rm d} x} \psi_k(x,t) \right)} \nonumber \\
&=& 2 \int_{0}^{k_F} \frac{{\rm d} k}{2 \pi} \;\;
{\rm Im}\, \left( \psi_k^* \frac{{\rm d}}{{\rm d} x} \psi_k +
\psi_{-k}^* \frac{{\rm d}}{{\rm d} x} \psi_{-k} \right)
\label{currdens}
\end{eqnarray}
where the prefactor $2$ comes from spin and $k_F=\sqrt{2 \ve_F}$ is the 
Fermi wavevector. 

The numerical parameters are as follows: the Fermi energy is 
$\ve_F=0.3$ a.u., the bias is $U_L=-U_R=0.05,0.15,0.25$ a.u., the central region 
extends from $x=-6$ to $x=+6$ a.u. with equidistant grid points with spacing 
$\Delta x=0.03$ a.u.. The $k$-integral in Eq.~(\ref{currdens}) is discretized 
with 100 $k$-points which amounts to a propagation 
of 200 states. The time step for the propagation was 
$\Delta t= 10^{-2}$ a.u.

In Fig.~\ref{examp1} we have plotted the current densities at $x=0$ as a 
function of time for different values of the applied bias. 
As a first feature we notice that a steady state is achieved, 
in agreement with the results of Ref. \onlinecite{saprb04}. 
The corresponding steady-state current $I$ can be calculated from the 
Landauer formula. For the present geometry this leads to the steady current 
\begin{widetext}
\begin{equation}
I=8e\int_{\max(U_{L},U_{R})}\frac{d\w}{2\p}[f(\w-U_{L})-f(\w-U_{R})]
\frac{\sqrt{\w-U_{L}}\sqrt{\w-U_{R}}}
{\left[\sqrt{\w-U_{L}}+\sqrt{\w-U_{R}}\right]^{2}+U_{L}U_{R}
\left[\frac{\sin(l\sqrt{2\w})}{\sqrt{\w}}\right]^{2}},
\label{land}
\end{equation}
\end{widetext}
where $l$ is the width of the central region. 
From Eq. (\ref{land}) with $l=12$ a.u. and $U_L=-U_R$, the 
numerical values for the steady-state currents  
are $0.0316$ a.u. ($U_{L}=0.05$ a.u.), $0.0883$ a.u. ($U_{L}=0.15$ a.u.) and 
$0.0828$ a.u. ($U_{L}=0.25$ a.u.). 
We see that our algorithm yields the same answers. Second, 
we notice that the onset of the current is delayed in relation to the 
switching time $t=0$. This is easily explained by the fact that the 
perturbation at $t=0$ happens in the leads only, {\em e.g.}, for  
$|x|> 6$ a.u., 
while we plot the current at $x=0$. In other words, we see the delay time 
needed for the perturbation to propagate from the leads to the center of 
our device region. We also note that the higher the bias the more the 
current overshoots its steady-state value for small times after switching 
on the bias. Finally it is worth mentioning that increasing the bias 
not necessarily leads to a larger steady-state current. 

In the second example we studied a double square potential barrier with 
electrostatic potential $V(x) = 0.5$ a.u. for $5$ a.u. $\leq |x| \leq 6$ a.u. 
and zero otherwise. This time we switch on a constant bias in the left lead 
only, {\em i.e.}, $U_R=0$. The Fermi energy for the initial state is 
$\ve_F=0.3$ a.u.. The central region extends from $x=-6$ to $x=+6$ a.u. with 
a lattice spacing of $\Delta x=0.03$ a.u.. Again, we use 100 different 
$k$-values to compute the current and a time step of 
$\Delta t = 10^{-2}$ a.u.. 

\begin{figure}[t]
\includegraphics[width=0.9\columnwidth]{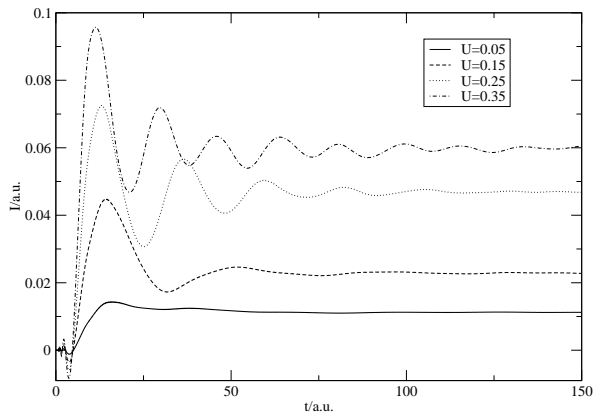}
\caption{\label{examp2} Time evolution of the current through a double 
square potential barrier in response to an applied constant bias (given in 
atomic units) in the left 
lead. The potential is given by $V(x) = 0.5$ a.u. for $5 \leq |x|\leq 6$ a.u. 
and zero otherwise, the propagation region extends from $x=-6$ to $x=+6$ a.u.. 
The Fermi energy of the initial state is $\ve_F=0.3$ a.u.. The current in the 
center of the structure is shown.}
\end{figure}

In Fig.~\ref{examp2} we plot the current at 
$x=0$ as a function of time for several values of the applied bias $U=U_L$. 
Again, a steady state is achieved for all values of $U$. 
As discussed in Fig.~\ref{examp1} the transient current can exceed the
steady current; the higher the applied voltage the larger is this excess 
current and the shorter is the time when it reaches its maximum.
Furthermore, the oscillatory evolution towards the steady current solution
depends on the bias. For higher bias the frequency of the transient 
oscillations increases. For small bias the electrons at the bottom of the  
band are not disturbed and the transient process is 
exponentially short. On the other hand, for a bias close to the Fermi 
energy the transient process decays as a power law, due to 
the band edge singularity. As pointed out in Ref. \onlinecite{saprb04}, for 
non-interacting electrons the steady-state current develops by means 
of a pure dephasing mechanism. In our examples the transient process occurs 
in a femtosecond time-scale, which is much shorter than the  
relaxation time due to electron-phonon interactions. 

\begin{figure}[t]
\includegraphics[width=0.9\columnwidth]{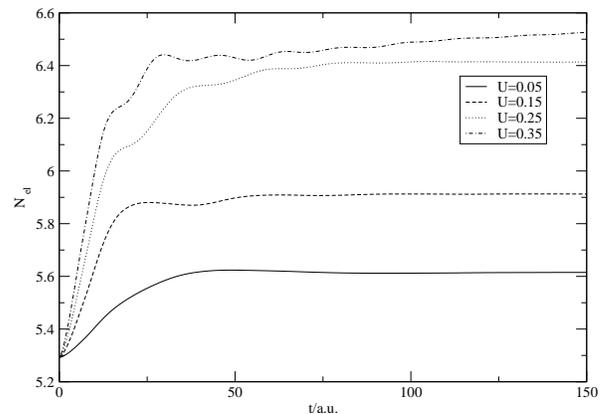}
\caption{\label{examp2norm} Time evolution of the total number of electrons 
in the region $|x|\leq 6$ for the double square potential barrier of 
Fig.~\ref{examp2}. }
\end{figure}

In Fig.~\ref{examp2norm} we plot the time evolution of the total number of 
electrons in the device region for the same values of $U_L$. We see that as 
a result of the bias a quite substantial amount of charge is added to the 
device region. This result has important implications when simulating
the transport through an interacting system as the effective (dynamical) 
electronic screening is modified due not only to the external field but also
to the accumulation of charge state in the molecular device. This illustrates 
that linear response might not be an appropriate tool to tackle the dynamical
response and that we will need to resort to a full time-propagation
approach as the one of the present work. 
Here we emphasize that all our calculations are done 
without taking into account the electron-electron interaction. If we had 
done a similar calculation with the interaction incorporated in a 
time-dependent Hartree or time-dependent DFT framework we would expect the 
amount of excess charge to be reduced significantly as compared to 
Fig.~\ref{examp2norm}.

\begin{figure}[t]
\includegraphics[width=0.9\columnwidth]{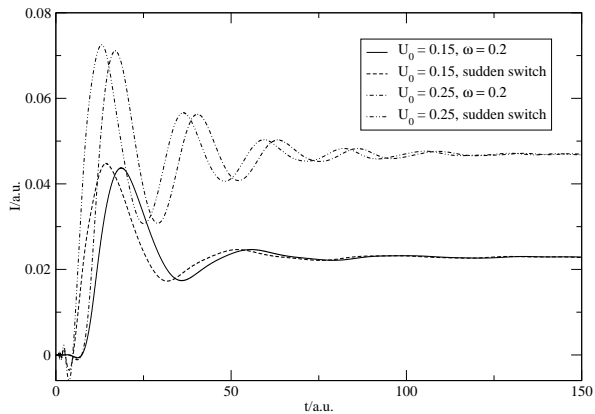}
\caption{\label{examp2a} Time evolution of the current for a double 
square potential barrier when the bias is switched on in two different 
manners: in one case, the bias $U_0$ is suddenly switched on at $t=0$ 
while in the other case the same bias is achieved with a smooth switching
$U(t) = U_0 \sin^2(\omega t)$ for $0<t<\pi/(2 \omega)$. The parameters 
for the double barrier and the other numerical parameters are the same as 
the ones used in Fig.~\ref{examp2}. $U_0$ and $\omega$ given in atomic units.}
\end{figure}

In Fig.~\ref{examp2a} we show time-dependent currents for the same double 
barrier as in Fig.~\ref{examp2} for two different ways of applying the 
bias in the left lead: in one case the constant bias $U_0$ is switched 
on suddenly at $t=0$ (as in Fig.~\ref{examp2}), in the other case the 
constant $U_0$ is achieved with a smooth switching $U(t)=U_0 
\sin^2(\omega t)$ for $0<t<\pi/(2 \omega)$. As expected and in agreement 
with the results of Ref.~\onlinecite{saprb04}, the same steady state 
is achieved after the initial transient time. However, the transient 
current clearly depends on how the bias is switched on. 

\begin{figure}[t]
\includegraphics[width=0.9\columnwidth]{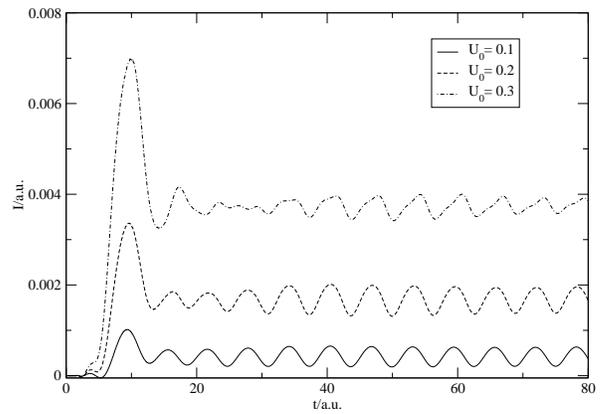}
\caption{\label{examp3} Time evolution of the current for a square potential 
barrier in response to a time-dependent, harmonic bias in the left lead, 
$U_L(t) = U_0 \sin(\omega t)$ for different amplitudes $U_0$ (values in a.u.) 
and frequency $\omega=1.0$ a.u.. The potential is given by $V(x)=0.6$ a.u. for 
$|x|\leq 6.0$ a.u. and zero otherwise. The propagation region extends from 
$x=-6$ to $x=+6$ a.u.. The Fermi energy of the initial state is 
$\ve_F=0.5$ a.u.. The current at $x=0$ is shown. }
\end{figure}

In the final example we address the simulation of ac-transport.
We computed the current for a single square potential 
barrier with $V(x) = 0.6$ for $|x|<6$ and zero otherwise. Here we 
applied a time-dependent bias of the form $U_L(t) = U_{0} \sin(\omega t)$ 
to the left lead. The right lead remains on zero bias. The numerical 
parameters are: Fermi energy $\ve_F=0.5$ a.u., device region from $x=-6$ to 
$x=+6$ a.u. with lattice spacing $\Delta x=0.03$ a.u.. The number of 
$k$-points is 100 and the time step is $\Delta t=  10^{-2}$ a.u..  
In Fig.~\ref{examp3} we plot the current at $x=0$ as a function of time for 
different values of the parameter $U_{0}=0.1,0.2,0.3$ a.u. The frequency was 
chosen as $\omega=1.0$ a.u. in both cases. Again, as for the dc-calculation  
discussed above, we get a transient that overshoots the average current 
flowing through the constriction; again, this excess current is larger the 
higher the applied voltage. Also, after the transient we obtain a current 
through the system with the same period as the applied bias. Note, however, 
that (especially for the large bias), the current is not a simple harmonic 
as the applied ac bias. 

\section{Summary and outlook}
\label{conc}

In the present work we have presented a formally
rigorous approach towards the description of charge transport using
an open-boundary scheme within TDDFT. We have implemented a 
specific time-propagation scheme that incorporates transparent 
boundaries at  the device/lead interface in a natural way.
In order to have a clear 
definition of a device region in Fig.~\ref{system} we assumed 
that an applied bias can be described by adding a spatially constant 
potential shift in the macroscopic part of the leads. This implies an 
effective ``metallic screening'' of the constriction. The screening limits 
the spatial extent of the induced density created by
the bias potential or the external field 
to the central region. Our treatment can be further refined to 
include dynamical screening deep inside the electrodes on the level of 
linear response, which might be of importance for the initial 
transient. 
Our time-dependent scheme allows to extract both ac and dc I/V 
device characteristics and it is ideally suited to describe
external field (photon) assisted processes.  

In order to illustrate the performance
of the method, we have implemented it for 
one-dimensional models and we have shown:
i) How to extract the proper initial extended states to be propagated. 
ii) How to incorporate perfect transparent boundaries for the time propagation.
iii) A steady-state current is always reached upon application
of a dc bias. The transient process occurs on a time-scale much 
shorter than the relaxation time due to electron-phonon interaction.
In the case of systems without any source of dissipation it 
is known that the steady-current is
independent of the history of the process.~\cite{saprb04}
We have explicitly demonstrated this history independence for two different 
switching processes of the external bias. However, if we allow for 
dissipation either through electron-electron or electron-phonon
interactions, the current versus voltage characteristics might depend
on the history. For instance, hysteresis loops due to different
transient electronic/geometrical device configurations are possible.
This effect will be more dramatic in the case of ac-driving fields of high 
frequencies where the system may not have enough time to respond to 
the perturbation.
iv) A periodic  ac current is reached upon perturbation with a
monochromatic field.

Previous work on time-dependent quantum transport mainly uses the idea of 
Caroli.\cite{caroli1,caroli2}
This approach is at the core of the Landauer derivation and has the problem
of using different chemical potentials in different parts of the system. 
This implies that the initial state is not a ground 
state of the entire, contacted system. Furthermore, the 
time-dependent perturbation is a tunneling Hamiltonian which is nonlocal 
in space. Therefore, it cannot be combined with TDDFT since 
there the time-dependent potential is local.

Here, we have used the scheme of Cini \cite{cini} which can be combined 
with TDDFT: We start the calculation from a well-defined thermodynamic
equilibrium configuration, therefore the scheme is thermodynamically 
consistent. Then we apply an external potential that in general is time 
dependent. By virtue 
of the Runge-Gross theorem, the time-evolution of this quantum system is 
completely determined by the knowledge of the time-dependent density. 
In the steady state regime the occupation of left and right moving carriers 
is dictated by the time-dependent Schr\"odinger equation. 

Another thermodynamically consistent scheme has been put forward by 
Kamenev and Kohn. \cite{kamenev:01} They used 
the microscopic quantum-kinetic formulation of conductivity 
and worked with the Kubo formula~\cite{kubo} 
in the linear time-dependent Hartree regime. 
They consider a closed system (ring) where the current in the system results 
from an external driving vector potential. This approach also overcomes
the problem of having two chemical potentials. They have shown that 
it is possible to recover the Landauer result, {\em i.e.}, the universal quantum 
of conductance $2e^2/h$ independent of the length and material. 
Since the Kamenev-Kohn approach uses a vector potential rather than a 
scalar potential, time-dependent current DFT rather than TDDFT would be the 
natural density-functional extension.

Most theoretical approaches to transport in molecular electronic devices adopt
open boundary conditions and assume that transport is ballistic, 
{\em i.e.}, under steady state conditions inelastic collisions are absent from 
a molecular structure and its 
contacts.~\cite{Landauer:57,Buettiker:86,book,Datta:95} 
Dissipation occurs only in the idealized macroscopic reservoirs 
connected by leads to the molecular device. This central role of the reservoirs
in the process of dissipation is a valid approximation, particularly 
when the applied bias is small and a device operates in a linear regime.
When inelastic scattering dominates this picture is not applicable.
In particular, experiments \cite{kushmerick,smit} indicate that inelastic 
scattering with lattice vibrations is present at sufficiently large bias, 
causing local heating of contacts and molecular devices. 
Energy transfer to the lattice may also cause atomic migration and 
result in dramatic changes in the device characteristics (also
they may give phonon replica structure in the measured conductance). 
The modelling of a many-electron system out of equilibrium coupled 
to lattice vibrations is a real theoretical challenge.~\cite{car} 

Electron correlations are also important in molecular conductors,
for example, Coulomb blockade effects dominate the transport in quantum dots.
Short-range electron correlations seems to be relevant in order to get
quantitative description of I/V characteristics in molecular
constrictions.~\cite{delaney:04,everts:04,ferreti}
In particular it is commonly assumed that the energy scales for 
electron-electron and electron-phonon interactions are different and
could be treated separately. However, the metallic screening of the
electrodes considerably reduces the Coulomb-charging energy (from eV to
meV scale). In this regime the energy scale for the two
interactions merge and they need to be treated on the same footing
posing some additional theoretical challenge.

Other approaches put forward in the literature directly look
for a homogeneous current-carrying state either based
on a a maximum entropy principle \cite{godby:03} or by a imposing
the current through Lagrange-multipliers.\cite{kosov:04} In those
approaches it is implicitly assumed
 that the origin of the homogeneous current is
independent whether it is introduced by reservoirs or by
external fields. This is indeed the case for independent electrons but once
dissipation is built in the system might exhibit a dependence on the history
of the applied bias,({\em e.g.}, possible appearance of
hysteresis loop in the current versus voltage characteristics).

It is clear that the quality of the TDDFT functionals is of crucial 
importance. In particular, exchange and correlation functionals for 
the non-equilibrium situation are required. 
Time-dependent linear response theory for dc-steady state 
has been implemented in Ref.~\onlinecite{baer:04} 
within TDLDA assuming jellium-like electrodes (mimicked by complex
absorbing/emitting potentials). It has been shown that the
dc-conductance changes considerably from the standard Landauer value.
Therefore, a systematic study of the TDDFT 
functionals themselves is needed. A step beyond
standard adiabatic approximations and exchange-only potentials
is to resort to many-body schemes as recently 
done for the characterization of optical properties of semiconductors and
insulators.~\cite{exc} Another path is to explore in depth 
the fact that the true exchange-correlation potential is 
current dependent.~\cite{vignale,tokatly}

An appealing feature of the present approach is that electron-electron
and electron-ion correlations and dissipation would in principle be described
correctly in two-component TDDFT.~\cite{kreibig} 

At present we are implementing our propagation scheme 
for real 3D systems~\cite{wirtz}  within the real-space real-time TDDFT 
code, {\em octopus}.~\cite{octopus} We are also
exploring the possibility of a semiclassical description of the 
electron-ion coupling in order to avoid the complexity of multicomponent
DFT and the problems related to mixed quantum classical approaches 
({\em i.e.}, Ehrenfest dynamics) as they fail to describe the long-term
inelastic electron-phonon scattering correctly.~\cite{alberto,todorov,car}

\section*{Acknowledgments}
This work was supported by the European Community 6th framework Network of
Excellence NANOQUANTA (NMP4-CT-2004-500198) and by the Research and Training 
Network EXCITING. AR acknowledges support from
the EC project M-DNA (IST-2001-38051), 
Spanish MCyT and the University of the Basque Country. We have benefited
from enlightening discussions with  L. Wirtz, A. Castro, H. Appel, M. 
A. L. Marques, C. Verdozzi and U. von Barth.

\appendix*

\section{Continued matrix fractions}
\label{app}

Let us consider the infinite tridiagonal block matrix 
\begin{equation}
M_{0}=\left[
\begin{array}{cccc}
A_{0} & B_{1}  & 0   & \ldots \\
B_{1} & A_{1} & B_{2} &   \ldots \\
0 & B_{2} & A_{2}   & \ldots \\
\ldots & \ldots & \ldots & \ldots 
\end{array}    
\right],
\end{equation}
where $A_{i}$ and $B_{i}$ are $N\times N$ matrices 
(the argument works even for matrices of different matching 
dimensions). We write the inverse matrix $M_{0}^{-1}$ as 
\begin{equation}
M_{0}^{-1}=\left[
\begin{array}{cc}
Q_{00} & \tilde{Q}_{01} \\ 
\tilde{Q}_{10} & Q_{1}  
\end{array}    
\right],
\end{equation}
where $Q_{00}$ is the first $N\times N$ block of $M_{0}^{-1}$. 
It is now convenient to introduce the matrix $M_{n}$ 
obtained by dropping the first $nN$ lines and $nN$ columns of 
$M_{0}$. Then, in terms of the rectangular matrix $\tilde{B}_{n}=
[B_{n},0,0,\ldots]$, we have
\begin{eqnarray}
\frac{1}{\left[
\begin{array}{cc}
    A_{0} & \tilde{B}_{1} \\
    \tilde{B}^{T}_{1} & M_{1}
\end{array}\right]}&=&
\frac{1}{\left[
\begin{array}{cc}
    A_{0} & 0 \\
    0 & M_{1}
\end{array}\right]}
 \\ &&-
\frac{1}{\left[
\begin{array}{cc}
    A_{0} & 0 \\
    0 & M_{1}
\end{array}\right]}\left[
\begin{array}{cc}
    0 & \tilde{B}_{1} \\
    \tilde{B}_{1}^{T} & 0
\end{array}\right]\frac{1}{\left[
\begin{array}{cc}
    A_{0} & \tilde{B}_{1} \\
    \tilde{B}^{T}_{1} & M_{1}
\end{array}\right]}.
\nonumber
\end{eqnarray}
From the above Dyson-like equation it is straightforward to obtain
$Q_{00}=A_{0}^{-1}-A_{0}^{-1}\tilde{B}_{1}\tilde{Q}_{10}$ and 
$\tilde{Q}_{10}=-M_{1}^{-1}\tilde{B}_{1}^{T}Q_{00}$. 
Solving for $Q_{00}$
\begin{equation}
Q_{00}=\frac{1}{A_{0}-\tilde{B}_{1}M_{1}^{-1}\tilde{B}_{1}^{T}}.
\end{equation}
One can now proceed along similar lines. We define
\begin{equation}
M_{1}^{-1}=\left[
\begin{array}{cc}
Q_{11} & \tilde{Q}_{12} \\ 
\tilde{Q}_{21} & Q_{2}  
\end{array}    
\right]
\end{equation}
where $Q_{11}$ is the first $N\times N$ block of $M_{1}^{-1}$. 
From the corresponding Dyson equation one finds
$Q_{11}=A_{1}^{-1}-A_{1}^{-1}\tilde{B}_{2}\tilde{Q}_{21}$ and 
$\tilde{Q}_{21}=-M_{2}^{-1}\tilde{B}_{2}^{T}Q_{11}$. 
Solving for $Q_{11}$
\begin{equation}
Q_{11}=\frac{1}{A_{1}-\tilde{B}_{2}M_{2}^{-1}\tilde{B}_{2}^{T}}
\end{equation}
and substituting this result back in $Q_{00}$ yields
\begin{equation}
Q_{00}=\mbox{$\frac{1}
{A_{0}-B_{1} \mbox{$\frac{1}
{A_{1}-\tilde{B}_{2} M_{2}^{-1} \tilde{B}_{2}^{T}}$}B_{1}}$}.
\end{equation}
Repeating the same steps we end up with the continued matrix fraction
\begin{equation}
Q_{00}=\mbox{$\frac{1}{A_{0}-B_{1}
\mbox{$
\frac{1}{A_{1}-B_{2}
\mbox{$\frac{1}{A_{2}-B_{3}
\mbox{$\frac{1}{A_{3}-B_{4}
\mbox{$\frac{1}{A_{4}\ldots}$}
B_{4}}$}
B_{3}}$}
B_{2}}$}
B_{1}}$}.
\end{equation}

\end{document}